\documentclass[11pt,a4paper]{article}
\usepackage{mcite}
\usepackage{epsfig}
\usepackage{axodraw}

\epsfverbosetrue

\def\3{\ss}

\newcommand{\tev}{{\rm Te}\kern-1.pt{\rm V}}
\newcommand{\gev}{{\rm Ge}\kern-1.pt{\rm V}}
\newcommand{\mev}{{\rm Me}\kern-1.pt{\rm V}}
\newcommand{\kev}{{\rm Ke}\kern-1.pt{\rm V}}
\newcommand{\gevsq}{\mbox{$\mathrm{{\rm Ge}\kern-1.pt{\rm V}}^2$}}
\newcommand{\gevmsq}{\mbox{$\mathrm{{\rm Ge}\kern-1.pt{\rm V}}^{-2}$}}

\newcommand{\mayor} {\mbox{\raisebox{-0.4ex}
{$\;\stackrel{>}{\scriptstyle \sim}\;$}}}

\newcommand{\sla}[1]{/\!\!\!#1}

\begin{document}
\begin{titlepage}
%\begin{flushright}
%~\\
%\end{flushright}

\begin{center}
\begin{huge}
\bf Search for Higgs Bosons Decay  $H\rightarrow \gamma\gamma$  Using Vector Boson Fusion \\
\end{huge}

\vspace{2.cm}

\Large Kyle Cranmer, Bruce Mellado, \\
\Large William Quayle, Sau Lan Wu \\
\vspace{0.5cm}
{\Large\it Physics Department \\
University of Wisconsin - Madison \\
   Madison, Wisconsin 53706 USA }
%\maketitle

\vspace{1.5cm}

\begin{abstract}
\noindent
The sensitivity of the ATLAS experiment to low mass SM Higgs produced via Vector Boson Fusion mechanism with $H\rightarrow \gamma\gamma$ is investigated. A cut based event selection has been chosen to optimize the expected signal significance with this decay mode. A signal significance of 2.2$\,\sigma$ may be achieved for $M_H=130\,\gev$ with 30 fb$^{-1}$ of accumulated luminosity. 
\end{abstract}
\end{center}
\setcounter{page}{0}
\thispagestyle{empty}

\end{titlepage}

\newpage

\pagenumbering{arabic}

\section{Introduction}
\label{sec:introduction}

In the Standard Model (SM), there are 4 gauge vector bosons
(gluon, photon, W and Z) and 12 fermions (six quarks and six
leptons)~\cite{np_22_579,*prl_19_1264,*sal_1968_bis,*pr_2_1285}.
These particles have been observed experimentally. The SM predicts
the existence of one scalar boson, the Higgs
boson~\cite{pl_12_132,*prl_13_508,*pr_145_1156,*prl_13_321,*prl_13_585,*pr_155_1554}.
The discovery of the Higgs boson remains one of the major
cornerstones  of the SM.

The observation of the Higgs boson is a primary focus of
the of ATLAS detector~\cite{LHCC99-14}. It is most interesting to
investigate the observability of the Higgs boson in the conditions of the LHC with the ATLAS detector.

\begin{figure}[ht]
{\centerline{\epsfig{figure=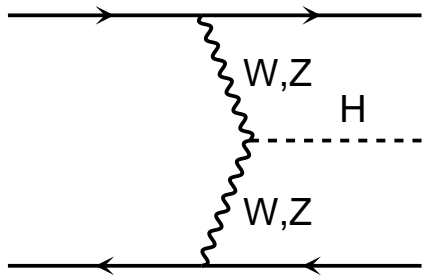,width=6.5cm}}}
\vspace{-0.5cm}
\caption[]{Lowest order Feynman diagram of the production of Higgs via VBF. The parallel solid lines correspond to quark lines.} \label{fig:vbf_feynman}
\end{figure}

The  Higgs at the LHC is produced predominantly via gluon-gluon
fusion. For Higgs masses, $M_H$,  such that $M_H>100\,\gev$, the
second dominant process is vector boson fusion (VBF). The lowest order
Feynman diagram of the production of Higgs via VBF is depicted in
Figure~\ref{fig:vbf_feynman}.

Early  analyses performed at the parton level with the decays
$H\rightarrow W^+W^-$  and $H\rightarrow \tau^+\tau^-$ via VBF indicated
that this mechanism could produce the most powerful discovery mode in the
range of the Higgs mass, $M_H$,
$115<M_H<200\,\gev$~\cite{pr_160_113004,*pl_503_113,*pr_61_093005}.
The ATLAS collaboration has performed feasibility studies for
these decay modes including more detailed detector description and
the implementation of initial state and final state parton showers
(IFSR), hadronization and multiple interactions~\cite{SN-ATLAS-2003-024}. Our group has contributed to this effort~\cite{ATL-PHYS-2003-002,*ATL-COM-PHYS-2003-002}.

In this note we consider the production of Higgs via VBF with $H\rightarrow\gamma\gamma$ . An early analysis performed at the parton level indicated that this process could be competitive with
the inclusive search~\cite{Rainwaterthesis}. Another analysis performed within the ATLAS Collaboration is available~\cite{Minagawa_040902,*Minagawa_121202,*Kanzaki_250603}.

The present feasibility study is addressed at low LHC luminosity  (10$^{33}$ cm$^{-2}$ s$^{-1}$) and the discovery potential is evaluated for an integrated  luminosity of 30 fb-1, which is expected to be accumulated during the first  years of LHC operation.

\section{MC Generation}
\label{sec:generation}

In this  Section details on the generation of signal and
background processes relevant to this analysis are given.

\subsection{Generation of Signal}
\label{sec:MCsignal}

The cross-section for the VBF process has been obtained with the
matrix element calculation provided within
PYTHIA6.1~\cite{cpc_82_74,*cpc_135_238}. The Higgs production
cross-sections with the VBF mechanism as a function of $M_H$ are given
in Table~\ref{tab:sigcross}. The Higgs branching ratio to two
$\gamma$'s  has been calculated with the package
HDECAY~\cite{cpc_108_56}. Table~\ref{tab:branching} shows the
values of the Higgs branching ratio to two $\gamma$'s as obtained
by PYTHIA6.1 and HDECAY programs. The values obtained with PYTHIA6.1 tend
to be larger  than those obtained with HDECAY. In the analysis we use the the branching ratio calculated with HDECAY.

\begin{table}[ht]
\begin{center}
\begin{tabular}{||c|c|c||}
\hline
  $M_H (\gev)$   & VBF  & g-g Fusion \\ \hline
110      & 4.65    &  22.12 \\ \hline
120              & 4.29    &  17.79 \\ \hline
130              & 3.97    &  16.17 \\ \hline
140              & 3.69    &  14.11 \\ \hline
150              & 3.45    &  12.50 \\ \hline
160              & 3.19    &  11.03 \\ \hline
 \end{tabular}
 \caption{Values of the Higgs production cross-section (in pb) with VBF and gluon-gluon fusion mechanisms  PYTHIA6.1 for different values of $M_H$.}
 \label{tab:sigcross}
\end{center}
\end{table}

\begin{table}[ht]
\begin{center}
\begin{tabular}{||c|c|c||}
\hline
  $M_H (\gev)$   & PYTHIA6.1  & HDECAY \\ \hline
110      & 1.868 10$^{-3}$  &  1.902 10$^{-3}$ \\ \hline
120      & 2.188 10$^{-3}$  &  2.186 10$^{-3}$ \\ \hline
130      & 2.297 10$^{-3}$  &  2.224 10$^{-3}$ \\ \hline
140      & 2.080 10$^{-3}$  &  1.934 10$^{-3}$ \\ \hline
150      & 1.543 10$^{-3}$  &  1.381 10$^{-3}$ \\ \hline
160      & 0.626 10$^{-3}$  &  0.537 10$^{-3}$ \\ \hline
 \end{tabular}
 \caption{Values of the Higgs branching ratio to two $\gamma$'s as obtained by PYTHIA6.1 and HDECAY for different values of $M_H$.}
 \label{tab:branching}
\end{center}
\end{table}

A sizable contribution from the production of Higgs via
gluon-gluon fusion is expected to appear. This note is concerned
with the feasibility of the observation of a Higgs signal with a
dedicated event selection meant to enhance the VBF signal. Therefore, the contribution from Higgs
production via gluon-gluon fusion is considered as a signal
processes. The production of this process has been modeled with
PYTHIA6.1. The Higgs production cross-sections with the gluon-gluon
fusion mechanism as a function of $M_H$ are given in
Table~\ref{tab:sigcross}.

\subsection{Generation of Background Processes}
\label{sec:MCbackground}

The relevant background processes are subdivided into two major
groups. Firstly, the production of two $\gamma$'s associated with
two jets. This will be called thereafter real photon production.
Secondly, a sizable contribution is expected from events in which
at least one jet is misidentified as a photon. This
background will be referred to as fake photon production. Despite
the impressive jet rejection rate after the application of
$\gamma$ selection criteria expected to be achieved by the ATLAS
detector~\cite{LHCC99-14} ($\mayor 10^3$ for each jet) the
contribution from fake photons will not be negligible due to the
large cross-sections of QCD processes at the LHC.

For the inclusion of hadronization, partonic showers and multiple
interaction effects the package PYTHIA6.2 is used. The
factorization and  renormalization scales are set to be equal. In
the case of $\gamma\gamma Nj$, $\gamma Nj$ and $Nj$ matrix elements (ME) for $N>0$ the
scales are set to the transverse momentum, $P_T$, of the lowest
$P_T$ parton. This choice of the scale will yield a rather conservative estimation of the cross-section specially for $\gamma Nj$ and $Nj$ processes. For the case of $\gamma\gamma$ ME the scales are set
to the invariant mass of the $\gamma$'s.

\subsubsection{Real Photon Production}
\label{sec:realphoton}

Several MC samples have been generated based on the $\gamma\gamma
jj$, $\gamma\gamma j$ and $\gamma\gamma$ matrix element
calculations. The $\gamma\gamma jj$ (QCD and EW~\footnote{Diagrams with $\gamma\gamma jj$ in the final state display four vertexes. A diagram is called QCD if at least a gluon appears in two of the vertexes. In EW diagrams no gluon appears in any of the vertexes.}) and  $\gamma\gamma
j$ ME based MC have been obtained from
MadGraphII~\cite{pc_81_357,*hep-ph_0208156,*MadGRAPHII}~\footnote{In
MadGraphII QCD and EW diagrams may be easily separated. EW
$\gamma\gamma jj$ diagrams are selected by setting the maximum QCD
order to 0. The QCD $\gamma\gamma jj$ ME code is obtained by
setting the maximum QCD and QED orders to 2.}.  The $\gamma\gamma$
ME based generator in PYTHIA6.1 has
been used. The latter contains the contribution from gluon-gluon
fusion via a quark loop, which is not available within MadGraphII.

The following cuts have been applied at the generator level:
\begin{itemize}
\item Minimum transverse momentum of the jets and photons is set to $20\,\gev$.
\item Pseudorapidity~\footnote{Pseudorapidity, $\eta$, is defined as $\eta=-\log({\tan{\theta/2}})$.} of the photons, $\eta_\gamma$, $\left|\eta_\gamma\right|<3$.
\item Pseudorapidity of the jets, $\eta_j$, $\left|\eta_j\right|<5$.
\item Distance in $R$~\footnote{$\Delta R$ is defined as $\sqrt{(\Delta\phi)^2+(\Delta\eta)^2}$.} between jets, $\Delta R_{jj}$, $\Delta R_{jj}>0.7$.
\item Distance in $R$ between jets and the $\gamma$'s, $\Delta R_{j\gamma}$, $\Delta R_{j\gamma}>0.3$.
\item Distance in $R$ between the $\gamma$'s, $\Delta R_{\gamma\gamma}$, $\Delta R_{\gamma\gamma}>0.1$.
\item Invariant mass of the $\gamma$ pair, $M_{\gamma\gamma}$, $80<M_{\gamma\gamma}<170\,\gev$.
\end{itemize}

In order to increase the efficiency of the generation of the QCD and
EW $\gamma\gamma jj$ ME MC samples used for analysis in
Sections~\ref{sec:partonlevel} and~\ref{sec:optimization}, a cut on
the difference in pseudorapidity between jets, $\Delta\eta_{jj}$,
is applied such that $\Delta\eta_{jj}>3$. The samples of QCD and
EW $\gamma\gamma jj$ events have been generated in intervals of
$M_{\gamma\gamma}$. The cross-sections obtained with QCD and EW
$\gamma\gamma jj$ ME based MC are given in Table~\ref{tab:aajjcross}.
A sample of QCD $\gamma\gamma jj$ ME MC used to perform studies reported  in Section~\ref{sec:doublecounting} has been generated without this additional cut on $\Delta\eta_{jj}$.

 \begin{table}[ht]
 \begin{center}
 \begin{tabular}{||c|c|c||}
 \hline
$M_{\gamma\gamma}$ Range ($\gev$) & QCD $\gamma\gamma jj$ & EW $\gamma\gamma jj$ \\
\hline\hline
 $80<M_{\gamma\gamma}<110$  &   1419    & 29.79  \\ \hline
 $110<M_{\gamma\gamma}<130$ &   581.4   & 16.14  \\ \hline
 $130<M_{\gamma\gamma}<170$ &   709.7   & 22.16  \\ \hline \hline
 $80<M_{\gamma\gamma}<170$  &   2710    & 68.09 \\ \hline

 \end{tabular}
 \caption{Cross-sections of QCD and EW $\gamma\gamma jj$ (in fb) for different ranges of $M_{\gamma\gamma}$  as calculated by MadGraphII. A cut on $\Delta\eta_{jj}>3$ has been applied on top of the cuts imposed at the generator level (see text).}
 \label{tab:aajjcross}
 \end{center}
 \end{table}

The contribution from double parton scattering (DPS), with pairs
of jets and photons coming from two independent parton collisions
is not considered in the final analysis. This process
contributed to some $10-15\,\%$ of the total background
in~\cite{Rainwaterthesis}. In addition to the background processes
studied in~\cite{Rainwaterthesis}  we consider the production of
fake photons. This background is a significant one (see
Section~\ref{sec:results}). Hence the relative contribution to the
total background from DPS will be about $5-7\,\%$.

\subsubsection{Fake Photon Production}
\label{sec:fakephoton}

The rate of fake photon production has been
estimated by generating samples with $\gamma jjj$ and $jjjj$ ME
based MC's. For this purpose MadGraphII is implemented.

This type of cross-section calculation involves thousands of diagrams. Generally speaking, it is convenient to separate EW and QCD diagrams. This speeds up the cross-section calculation~\footnote{In MadGraphII the script ``survey'' is called before event generation. The latter performs a quick integration over the phase space with various levels of optimization. These are meant to speed up the  ultimate event generation. However, in the presence of electro-weak bosons in internal lines more advanced levels of optimization may result into numerical instabilities. Therefore, the generation of EW processes is rather time consuming, specially when it comes to $2\rightarrow 4$ processes, as in this particular case.}. Unlike in the case of the $\gamma\gamma jj$ process, the EW $\gamma jjj$ diagrams are expected to contribute little. After the application of the cuts at the generator level used in the previous Section~\footnote{Generator cuts specified in the bullets.} (except for the cut on the invariant mass of the $\gamma\gamma$ pair) the QCD and EW $\gamma jjj$ diagrams yield 17.8 nb and 4.93 pb, respectively~\footnote{In QCD $\gamma jjj$ diagrams the dominant subprocesses contain at least one gluon in the initial state. These type of subprocesses are suppressed in the EW $\gamma jjj$ diagrams}. Further cuts have been applied at the generator level:
\begin{itemize}
\item Maximum invariant mass between the $\gamma$ and the jets (or between the jets in the case of $jjjj$ ME) should be at least $100\,\gev$.
\item The maximum difference in pseudorapidity between jets is required to be at least 3.5 units.
\end{itemize}
After the application of these additional cuts the QCD and EW $\gamma jjj$ diagrams produce 6.32 nb and 1.21 pb, respectively. Assuming an effective jet rejection of the order of $10^{3}$, the starting cross-section for the EW $\gamma jjj$ process would be $\approx 1\,$fb. This small cross-section will be severely reduced after the application of further selection cuts (see Section~\ref{sec:optimization}). In the physics analysis EW $\gamma jjj$ diagrams will be neglected. From now on the $\gamma jjj$ ME MC will include  QCD diagrams only.

The situation with the $jjjj$ process is similar. Only QCD $jjjj$ diagrams will be considered in the analysis. After the application of the cuts at the generator used in the previous section a cross-section of 24650 nb is obtained. The enhancement of the $jjjj$ cross-section over that of the $\gamma jjj$ is striking, being at least two orders of magnitude greater than the ratio of QCD to QED coupling constants. The main contributors to the cross-sections are the subprocesses with at least one gluon in the initial state and at least two gluons in the final state. Apart from the appearance of purely gluonic diagrams~\footnote{As a matter of fact, the subprocess $gg\rightarrow gggg$ takes up 45$\%$ of the cross-section.} a number of diagrams in subprocesses with a quark in the initial and final state, $qg\rightarrow qggg$, appear such that the gluons in the final state come from gluon splitting.

In order to pin down severe divergence effects the cross-section from the purely gluonic subprocesses $gg\rightarrow gggg$ and $gg\rightarrow ggg$ are compared at a fixed scale (the mass of the Z boson). The cross-section  for $gg\rightarrow ggg$ is $\approx 6$ times larger than that of the $gg\rightarrow gggg$ subprocess. This is consistent with moderate divergence effects.

Despite the large cross-section for the $jjjj$ ME the contribution of this process to the VBF analysis is not expected to overwhelm the total background contribution. The transverse momentum distribution of the lowest $P_T$ jets (the jets that are most likely to turn into a fake photon) with this process falls extremely rapidly. Harder cuts on the $P_T$ of the jets and photons in the VBF analysis will significantly reduce the contribution from this background (see Section~\ref{sec:results}).

The estimation of the fake photon background based on $\gamma j$
and $jj$ ME MC is not used here for the final results (see
Section~\ref{sec:results}). In this case one or two tagging jets
would come from the parton shower. Detailed studies performed on
the production of the $Z$ boson associated with two well separated
jets have shown that the rate and the angular correlations between
the tagging jets and the decay products of the boson are not
described well when at least one tagging jet is produced by the
parton shower~\cite{ATL-COM-PHYS-2003-042,*Mellado_12_12_02}. The deviation from
the full ME description goes beyond leading order (LO)
uncertainties and it is strongly dependent on the scale set to the
IFSR parton showers. It may be anticipated that the rate of fake
$\gamma\gamma$ associated with two well separated jets obtained
with the $\gamma j$ and $jj$ ME based MC  will severely
underestimate a more reliable rate obtained with the $\gamma jjj$
ME based MC.

In order to test these assumptions a sample of $\gamma j$ ME based
MC is produced with PYTHIA6.2. The rates of $\gamma\gamma$
associated with two well separated jets will be given in
Section~\ref{sec:optimization} and compared to those obtained on
the basis of the $\gamma jjj$ ME.

\section{Parton Level Analysis}
\label{sec:partonlevel}

As a first step, a parton level analysis is performed without the
inclusion of parton shower, hadronization and multiple interaction
effects. This will allow a direct comparison with the parton level
analysis performed in~\cite{Rainwaterthesis}. In the latter work
an earlier version of  MadGraph was used to generate the signal
and background MC samples. There one signal process was
considered and the fake photon background was not considered.

The following event selection adopted in~\cite{Rainwaterthesis} is
used here:
\begin{itemize}
\item [{\bf a.}] Minimum transverse momentum  of the $\gamma$'s,
$P_{T\gamma 1}>50\,\gev$ and $P_{T\gamma 2}>25\,\gev$. Here
$P_{T\gamma 1}$ and $P_{T\gamma 2}$ correspond to the $P_T$ of the
first highest and second highest transverse momentum $\gamma$'s,
respectively. The $\gamma$'s are required to fall in the central
region of the detector ($\left|\eta\right|<2.5$).
\item [{\bf b.}] The presence of two tagging jets~\footnote{Tagging jet candidates are defined as the two highest $P_T$ jets in the event.} in opposite
hemispheres is required. The tagging jets are required to lie within the
acceptance of the detector ($\left|\eta\right|<5$). The $P_T$ of
the leading jet, $P_{T1}$, should be $P_{T1}>40\,\gev$. The $P_T$
of the second highest $P_T$ jet, $P_{T2}$, is required to be
$P_{T2}>20\,\gev$. The tagging jets should be well separated, with
$\Delta\eta_{jj}>4.4$.
\item [{\bf c.}] The $\gamma$'s should be in pseudorapidity in between the tagging
jets with a buffer of $0.7$ units.
\item [{\bf d.}] No explicit requirement on the invariant mass of the
tagging jets is applied.
\item [{\bf e.}] No central jet veto survival probability correction is applied~\footnote{Cuts {\bf d} and {\bf e} are applied in the final
event selection. These bullets are placed here in order to avoid confusion
in Sections~\ref{sec:optimization} and~\ref{sec:results}.}.
\item [{\bf f.}] The invariant mass of the $\gamma$'s  should be $M_H-1<M_{\gamma\gamma}<M_H+1\,\gev$.

\end{itemize}

The experimental photon finding efficiency was chosen to be
$80\,\%$. The efficiency of matching a parton to a jet was set to
$86\,\%$ independent on the pseudorapidity. Thus, the combined
detector efficiency associated to each event is $0.473$. The
photon finding efficiency correction is applied after cut {\bf a}.
The parton-jet matching efficiency has been applied here after cut
{\bf b}.

 \begin{table}[ht]
 \begin{center}
 \begin{tabular}{||c|c|c|c|c||c||}
 \hline
 & {\bf a}    & {\bf b} & {\bf c}  & {\bf f} &  \cite{Rainwaterthesis} \\
\hline\hline
 VBF &  3.70   & 1.00    &  0.87   &  0.54  & 0.63 \\ \hline
 QCD $\gamma\gamma jj$ & 169.08  & 17.55  & 5.11  & 0.52 & 0.41\\ \hline
EW $\gamma\gamma jj$  & 5.23  & 1.93  &  1.53 & 0.15 & 0.16 \\ \hline
 \end{tabular}
 \caption{Effective cross-sections at parton level after successive cuts (see text). Cross-sections are given in fb for VBF signal ($M_H=120\,\gev$) and the real photon background,  QCD and EW $\gamma\gamma jj$. The last column corresponds to the results quoted in Rainwater's thesis (see text).}
 \label{tab:parton}
 \end{center}
 \end{table}

The final state particle four-momenta are passed through the
ATLFAST~\cite{ATLFAST} package. This includes the
smearing of the energy/momentum and position reconstructions. The
parameters of the smearing applied in~\cite{Rainwaterthesis} are
somewhat different. Additionally, here we use the proton structure
function CTEQ5L where in~\cite{Rainwaterthesis} CTEQ4L was used
instead.

The effective cross-sections after successive cuts for VBF signal
($M_H=120\,\gev$), QCD and EW $\gamma\gamma jj$ are shown in
Table~\ref{tab:parton}. The results obtained
in~\cite{Rainwaterthesis}  after all cuts, before the application
of the central jet veto survival probability are shown in the last
column. The VBF signal and the QCD $\gamma\gamma jj$ effective
cross-section obtained here are respectively $15\%$ smaller and $27\%$ larger than
those obtained in~\cite{Rainwaterthesis}~\footnote{A detailed analysis of the source of these discrepancies has not been performed here. However, good agreement was found between our group's results 
and~[8] with regards to signal, QCD and EW $\gamma\gamma jj$ production. The comparison of the contribution from fake photons is still ongoing.}.

\section{Double Counting in Real QCD $\gamma\gamma jj$ Background}
\label{sec:doublecounting}

In the present note the effect of initial and final state
radiation is included.  Events with $\gamma\gamma$ and two
additional jets may be generated with $\gamma\gamma j$ ME when the
second tagging jet comes from IFSR. Alternatively, two additional
jets may be generated with $\gamma\gamma$ ME when the two
tagging jets are produced in the parton shower. The question
arises whether the $\gamma\gamma jj$ background rate calculated
with the QCD $\gamma\gamma jj$ ME MC yields a conservative enough
estimation from the point of view of a LO analysis.

In order to study the interplay between the ME and IFSR based
production of two partons associated with $\gamma\gamma$ several
MC samples have been analyzed. The analysis is performed at the
parton level.  Partons in the final state are ordered
according to $P_T$. The ME generators are interfaced with PYTHIA6.2 in
order to perform IFSR. In order to obtain the four-momenta of the jet
originating from IFSR a clustering procedure is performed over the
partons resulting from the cascade (before any hadronization
occurs).

\begin{figure}[ht]
{\centerline{\epsfig{figure=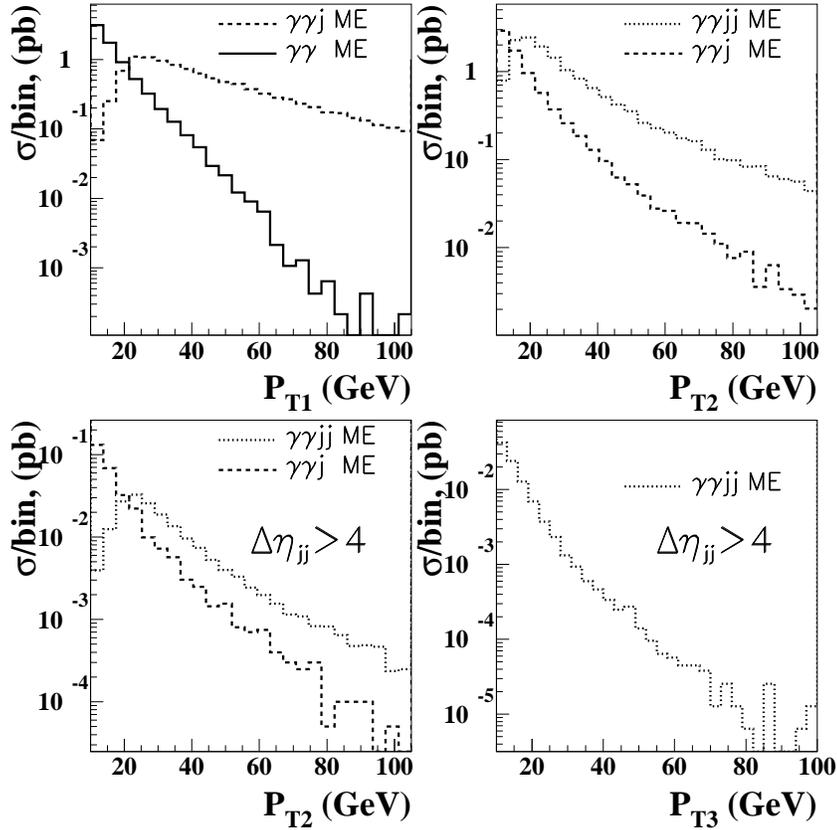,width=11.3cm}}}
\caption[]{Transverse momentum distributions of leading jets associated with $\gamma\gamma$ production. The upper left plot displays the $P_T$ of the leading jet produced by the $\gamma\gamma$ ME (solid line) and the $\gamma\gamma j$ ME. The upper right shows the $P_T$ of the second jet produced by the $\gamma\gamma j$ ME  (dashed line) and the QCD $\gamma\gamma jj$ ME (dotted line). The lower left plot shows similar distributions for events with $\Delta\eta_{jj}>4$. The lower right plot corresponds to the $P_T$ of the third jet obtained with QCD $\gamma\gamma jj$ ME for events with $\Delta\eta_{jj}>4$.} \label{fig:matching}
\end{figure}

The comparison between the ME and IFSR based production of additional
partons is made in two steps. 
%Firstly, the production rates of
%$\gamma\gamma$ plus the leading jet, $P_{T1}$ originating from the 
%$\gamma\gamma$ and $\gamma\gamma j$ ME are compared.
Figure~\ref{fig:matching} shows the $P_T$ distributions of the
leading jet obtained with the $\gamma\gamma$ ME (solid line) and
with the $\gamma\gamma j$ ME (dashed line). Here no additional
requirements are applied on $\Delta\eta_{jj}$ on top of the cuts
performed at the generator level (see Section~\ref{sec:realphoton}).

In the physics analysis the cut on $P_{T1}$ lies between 40 and
50$\,\gev$. For these values of $P_{T1}$ the differential
cross-section obtained with $\gamma\gamma j$ ME is well above the
one obtained with $\gamma\gamma$. This remains the case for events
with large rapidity gaps, $\Delta\eta_{jj}>4$. The $\gamma\gamma$
ME and $\gamma\gamma j $ ME curves may be matched by ``fudging''
the latter in order to meet the condition that the total
cross-section be consistent with the next-to-leading order (NLO)
cross-section. The K factor resulting from the NLO correction to
the non-resonant production of $\gamma\gamma$ is about a factor of two~\cite{Unal_12_12_02,pr_66_074018,epc_16_311}. 

%In order to
%achieve this the surface below the $\gamma\gamma j $ ME curve
%needs to be reduced by approximately a factor of
%two~\footnote{Alternatively, one can think of generating partons
%with $P_T$ as low as $10\,\gev$. The two curves may eventually
%cross at $\approx 10-15\,\gev$. Two curves would have to be
%lowered in order to match the NLO cross-section.}.

The upper right plot in Figure~\ref{fig:matching} displays the
$P_T$ of the second jet produced by the $\gamma\gamma j$ ME
(dashed line) and the QCD $\gamma\gamma jj$ ME (dotted line). The
QCD $\gamma\gamma jj$ ME curve always remains above the
$\gamma\gamma j$ ME curve. This remains true for events with
$\Delta\eta_{jj}>4$, as illustrated in the lower left plot of
Figure~\ref{fig:matching}.

For a LO type of analysis, the estimation of the $\gamma\gamma jj$
background obtained with the QCD $\gamma\gamma jj$ ME  MC~\footnote{In
addition, of course, to the EW $\gamma\gamma jj$ ME MC.} yields a
conservative enough estimation. The addition of contributions from
the $\gamma\gamma $ and $\gamma\gamma j$ ME based MC's will lead
to straight double counting, and, therefore, it will result into
an unnecessary overestimation of the $\gamma\gamma jj$ background.

%In order to reduce the theoretical uncertainty on the background
%normalization to the level closer to that of an  NLO calculation a
%more complex procedure should be followed by means a two step
%matching. Yet, a systematic error due to the different ways in
%which the matching between curves may be achieved needs to be
%computed.

It is relevant to note that the central jet veto survival
probability for QCD $\gamma\gamma jj$ calculated based on the
parton shower approach is significantly larger than that
calculated in~\cite{Rainwaterthesis}. The lower right plot in
Figure~\ref{fig:matching} shows the $P_T$ of the third jet,
$P_{T3}$, produced by the QCD $\gamma\gamma jj$ ME MC for events
with $\Delta\eta_{jj}>4$. The probability of having an additional
(non tagging) jet with $P_{T3}>20\,\gev$ in the central detector
region is $~25\%$~\footnote{This is pretty much independent on the
type of hard scattering, whether we deal here with a QCD or an EW
process (see Section~\ref{sec:results} and
Table~\ref{tab:cross120}).}. This leads to a jet veto survival
probability of the order of 75$\%$ for the QCD background, much
larger than the 30$\%$ calculated in~\cite{Rainwaterthesis}. This
results into an enhancement of the QCD $\gamma\gamma jj$
background with respect to the parton level based estimate
obtained in~\cite{Rainwaterthesis} by a factor of $\approx
2.5$~\footnote{Another factor of two arises from the different
size of the search mass window chosen in the physics analysis (see
Section~\ref{sec:optimization}).}.

\section{Detector Effects}
\label{sec:detector}

The smearing of the energy/momentum and position reconstructions
and jet clustering effects are performed with the help of the fast
simulation package ATLFAST~\cite{ATLFAST}. This package
provides a parametrized response of the detector, based on full GEANT simulation
results. The $M_{\gamma\gamma}$
resolution obtained with the help of the fast simulation is
$1.2\,\%$ for signal with $M_H=120\,\gev$.

 The energy
scale of the jets are corrected with the package
ATLFASTB~\cite{ATLFAST}.  The parton-jet matching efficiency and
the central jet fake veto obtained with ATLFAST are corrected with
the help of dedicated routines~\cite{ATL-COM-CAL-2002-003}. The
photon finding efficiency is assumed to be $80\%$.

The probability of a hadronic jet to be observed as a photon is
available in a study presented in the ATLAS
TDR~\cite{LHCC99-14}~\footnote{Volume I, page 223.}. This has been
parameterized as a function of the $P_T$ of the jet. The
parameterization of the central values of the fake photon
probability, $P_{fp}$, at low luminosity yields:
\begin{center}
\[
P_{fp}(P_T) = \left\{ \begin{array}{ll} 1/p_3(P_T) & 20<P_T<50\,\gev ;  \\
 1/3400 &  P_T>50\,\gev, \end{array} \right.
\]
\end{center}
where, $p_3$ is a third order polynomial with parameters,
$a_0=-3300, a_1=335.67, a_2=-6.45, a_3=0.04833$. The determination
of the fake photon probability is subject to systematic errors.
Large errors are due to the MC statistics which was
available for the initial study~\footnote{According to the ATLAS
TDR the jet rejection at low luminosity for $P_T=20\,\gev$ is
$1270\pm80$, for $P_T=40\,\gev$ is  $2900\pm300$. The error
increases with $P_T$.}. Additionally, the fake photon probability is process dependent. A study will be available in the near future, which will address these issues in more detail.

\section{Optimization of the Event Selection}
\label{sec:optimization}

In this Section an event selection is obtained by means of
maximizing the single bin Poisson significance for 30 fb$^{-1}$ of
accumulated luminosity and $M_H=120\,\gev$. The maximization
procedure is performed with the help of the MINUIT program. A
number of variables are chosen that are sensitive to the different
kinematics displayed by the signal and background processes. These
are common to the feasibility studies performed on most of the VBF
production modes~\footnote{For a detailed discussion
see~\cite{Rainwaterthesis}.}. The following variables are chosen:
\begin{itemize}
\item Transverse momentum of the tagging jets.
\item Difference in pseudorapidity and azimuthal angle between the tagging jets, $\Delta\eta_{jj}$ and $\Delta\phi_{jj}$, respectively.
\item Invariant mass of the tagging jets, $M_{jj}$.
\item Transverse momentum of the photons.
\item Difference in pseudorapidity and azimuthal angle between photons, $\Delta\eta_{\gamma\gamma}$ and $\Delta\phi_{\gamma\gamma}$, respectively.
\end{itemize}

Due to the implementation of parton shower and hadronization
effects the kinematics of the final state  will be somewhat
different from that  of the parton level analysis. In the
present analysis the contribution from fake photon production has
been included. As a result, the event selection needs to be
re-optimized~\footnote{It is worth noting that here we optimize
the Poissonian significance as opposed to the Gaussian
approximation, $S/\sqrt{B}$. The optimization is also sensitive to
this feature of the confidence level calculation~\cite{ATL-PHYS-2003-008}.}.

A number of pre-selection cuts are applied
similar to those used to obtain the multivariate optimization in
the VBF $H\rightarrow W^+W^- \rightarrow l^{+}l^{-}\sla{p_{T}}$
analysis~\cite{ATL-PHYS-2003-007}:
\begin{itemize}
\item [{\bf a.}] $P_{T\gamma 1}, P_{T\gamma 2}>25\,\gev$. The $\gamma$'s are required to fall in the central
region of the detector excluding the interface between the barrel and end-cap calorimeters ($1.37<\left|\eta_\gamma\right|<1.52$). The latter requirement reduces the acceptance by about 10$\%$.
\item [{\bf b.}] Two tagging jets in opposite hemispheres~\footnote{Tagging jets are defined as the two highest $P_T$ jets in the event.}, with $P_{T1}, P_{T 2}>20\,\gev$ and $\Delta\eta_{jj}>3.5$.
\item [{\bf c.}] The $\gamma$'s should be in pseudorapidity in between the tagging jets (no buffer is required).
\item [{\bf d.}] Invariant mass of the tagging jets, $M_{jj}>100\,\gev$.
\item [{\bf e.}] Central jet veto. No additional (non tagging) jets with $P_T>20\,\gev$ should be observed within $\left|\eta\right|<3.2$.
\item [{\bf f.}] The invariant mass of the $\gamma$'s should be $M_H-2<M_{\gamma\gamma}<M_H+2\,\gev$.
\end{itemize}

The photon finding efficiency correction is applied after cut {\bf
a}. The forward jet tagging efficiency and the fake central jet
veto rate corrections are applied after cuts {\bf b} and {\bf e},
respectively. Table~\ref{tab:preseleccross} shows the effective
cross-sections (in fb) for signal and background processes after
the application of cuts {\bf e} and {\bf f}. The dominant
background corresponds to the  QCD $\gamma\gamma jj$ and the fake
photon production, therefore, the optimization process will be
mainly determined by the kinematics of these process together with
that of the VBF signal.

\begin{table}[ht]
\begin{center}
\begin{tabular}{||c||c|c||c|c||c|c|c||}
\hline
  Cut   & VBF  & g-g Fusion & QCD $\gamma\gamma jj$ & EW $\gamma\gamma jj$ & $\gamma j$ & $\gamma jjj$  & $jjjj$ \\ \hline
{\bf e}  &   1.04 &  0.25   & 117.9 & 10.84 & 40.32 & 45.  & 109.57 \\ \hline
{\bf f}  &   0.94 &  0.22   & 5.67     &  0.52     & 0.68 & 4.19 &  10.24 \\ \hline
  \end{tabular}
 \caption{Effective cross-sections (in fb) for signal and background processes after the application of cuts {\bf e} and {\bf f}.}
 \label{tab:preseleccross}
\end{center}
\end{table}

The sixth column of Table~\ref{tab:preseleccross} shows the results
of fake photons obtained with the $\gamma j$ ME based MC. As
anticipated in Section~\ref{sec:fakephoton}, the rate of fake
$\gamma\gamma$ associated with two well separated jets predicted
by the $\gamma j$ ME is expected to undershoot that obtained with
the $\gamma jjj$ ME. Additionally, the jet and photon $P_T$
distributions are significantly steeper in the case of the $\gamma
j$ ME based MC. This will further suppress the contribution from
this MC in the optimized event selection.

\begin{figure}[ht]
{\centerline{\epsfig{figure=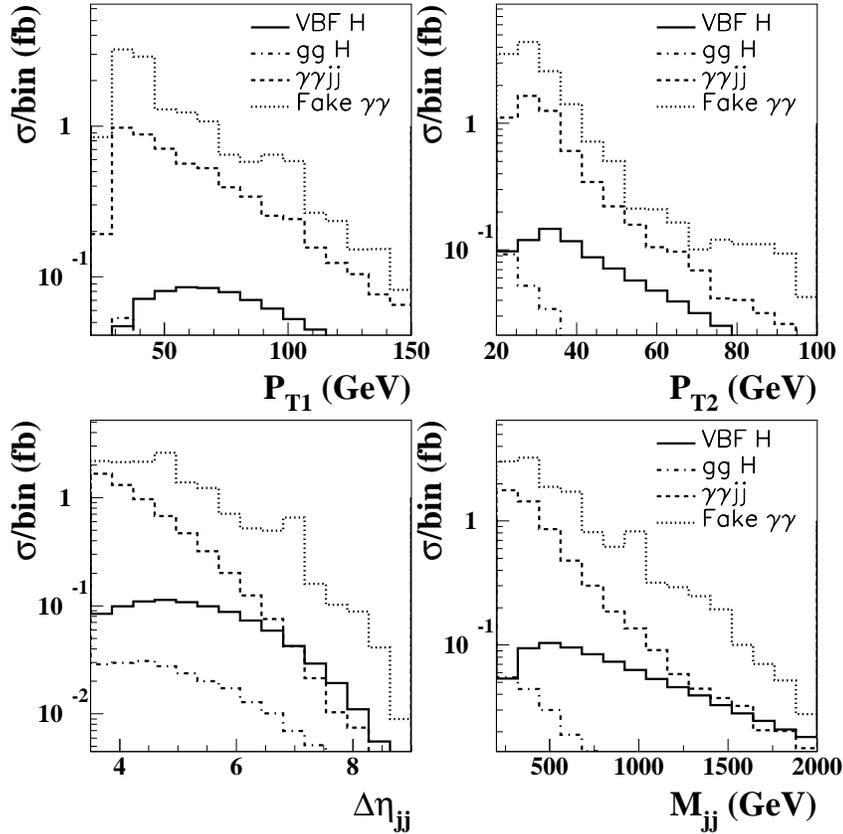,width=11.3cm}}}
\caption[]{Kinematic distributions of signal and background
processes after pre-selection cuts (see text). The upper left and
upper right plots correspond to the transverse momentum of the
leading jets. The lower left and lower right plots show the
difference in pseudorapidity between the
leading jets and their invariant mass, respectively. } \label{fig:vbfgg_1}
\end{figure}

\begin{figure}[ht]
{\centerline{\epsfig{figure=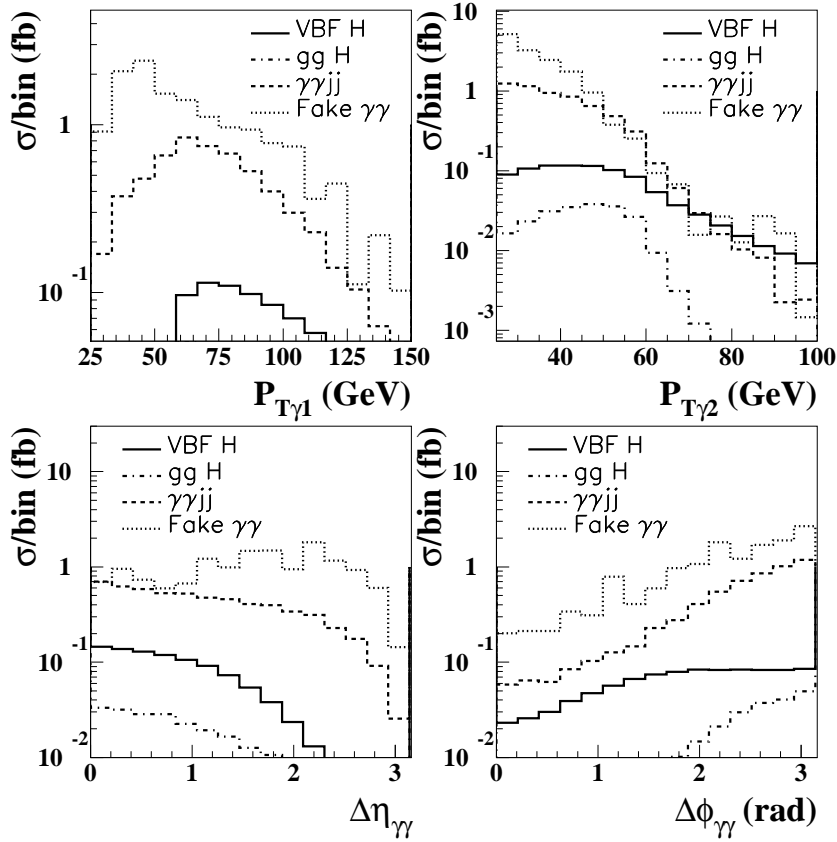,width=11.3cm}}}
\caption[]{Kinematic distributions of signal and background
processes after pre-selection cuts (see text). The upper
left  and right plots display the $P_T$ of the $\gamma$'s.
The lower left and right  plots show the difference in pseudorapidity
and in azimuthal angle between the  $\gamma$'s.} \label{fig:vbfgg_2}
\end{figure}

\begin{table}[ht]
\begin{center}
\begin{tabular}{||c|c|c|c||}
\hline
  Cut   & Pre-selection  & Parton Level & Optimization  \\ \hline
 {\bf a} &  $P_{T\gamma 1}, P_{T\gamma 2} >25\,\gev$  &     $P_{T\gamma 1}>50\,\gev$    &     $P_{T\gamma 1}>57\,\gev$   \\

& &  $P_{T\gamma 2}>25\,\gev$ &  $P_{T\gamma 2}>34\,\gev$ \\

& & & $\Delta\eta_{\gamma\gamma}<1.58$, $\Delta\phi_{\gamma\gamma}<3$~rad \\ \hline

 {\bf b} &   $P_{T1}, P_{T2} >20\,\gev$  &  $P_{T1}>40\,\gev$
 & $P_{T1}>40\,\gev$ \\

 & & $P_{T2}>20\,\gev$ & $P_{T2}>29.5\,\gev$ \\

& $\Delta\eta_{jj}>3.5$ &  $\Delta\eta_{jj}>4.4$ &  $\Delta\eta_{jj}>3.9$  \\ \hline

{\bf d} &   $M_{jj}>100\,\gev$    &     -  &     $M_{jj}>610\,\gev$ \\
\hline

  \end{tabular}
 \caption{Values of the cuts applied in the pre-selection and the optimized event selection compared to those applied for the parton level analysis (see Section~\ref{sec:partonlevel}).}
 \label{tab:cuts}
\end{center}
\end{table}

Figures~\ref{fig:vbfgg_1}-\ref{fig:vbfgg_2} display the
distributions of the variables chosen for the optimization of the
event selection after the application of pre-selection cuts.
The upper left and upper right plots in Figure~\ref{fig:vbfgg_1}
correspond to the transverse momentum of the leading jets. The
lower left and lower right plots in Figure~\ref{fig:vbfgg_1} show
the difference in pseudorapidity between the leading jets and
their invariant mass, respectively. The upper left  and right
plots in Figure~\ref{fig:vbfgg_2} display the $P_T$ of the
$\gamma$'s. The lower left and right  plots in
Figure~\ref{fig:vbfgg_2}  show the difference in pseudorapidity
and in azimuthal angle between the  $\gamma$'s.

Initially, it has been verified that the inclusion of additional
variables to those considered in~\cite{Rainwaterthesis} (see
Section~\ref{sec:partonlevel}) improves the signal significance.
The addition of the photon related variables,
$\Delta\eta_{\gamma\gamma}$ and $\Delta\phi_{\gamma\gamma}$,
improves the signal significance by some $10-20\%$ depending on
the Higgs mass. The implementation of those two variables
separately proves more efficient than the combined $\Delta
R_{\gamma\gamma}$. The inclusion of the hadronic variable
$\Delta\phi_{jj}$ does not noticeably increase the signal
significance. In the end  the optimization is performed with 8
variables excluding $\Delta\phi_{jj}$.

Table~\ref{tab:cuts} shows the results of the optimization
together with the values of the cuts placed at the pre-selection
level and for the parton level analysis.  Due to the significant
increase in the background contribution compared to the parton
level analysis~\footnote{The increase of the background comes from
the different choice of the width of the mass window, the
implementation of parton showers for the estimation of the central
jet veto probability (see Section~\ref{sec:doublecounting}) and the inclusion of fake photon events.} the
optimized event selection is significantly tighter, resulting into
reduced signal and background rates (see
Section~\ref{sec:results}).

\section{Results and Discovery Potential}
\label{sec:results}

Here we use  the event selection obtained in the optimization
procedure performed in Section~\ref{sec:optimization} (see
Table~\ref{tab:cuts}). The expected  signal and
background cross-sections corrected for acceptance and efficiency
corrections are shown in Table~\ref{tab:cross120}. Here the  mass
window is set for $M_H=120\,\gev$. In this table the results are
given after application of successive cuts. In
Table~\ref{tab:cross} results are given after the application of
all cuts.

The contribution from the fake photon background has been severely reduced thanks to the inclusion of the photon angular variables (see Figure~\ref{fig:vbfgg_2}).
The contribution from this background is, however, important. The
normalization of the fake photon background is subject to sizable
systematic uncertainties. This is due to the error on the
determination of the fake photon rejection rate (see
Section~\ref{sec:detector}).

\begin{figure}[ht]
{\centerline{\epsfig{figure=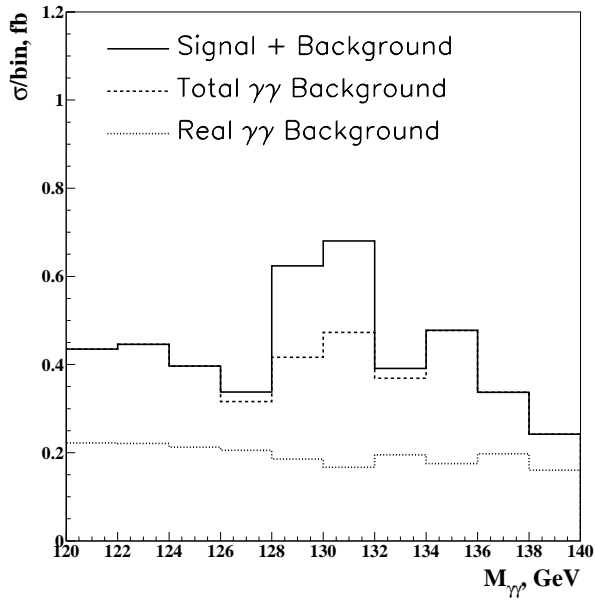,width=8.cm}}}
\caption[]{Expected signal and background effective cross-section in fb as a function of $M_{\gamma\gamma}$ for $M_H=130\,\gev$. The dashed line shows the total background contribution whereas the dotted line corresponds to the real $\gamma\gamma$ background. The solid line displays the expected contribution of signal plus background.} \label{fig:mgg}
\end{figure}

Figure~\ref{fig:mgg} shows the expected signal and background
effective cross-section in fb as a function of $M_{\gamma\gamma}$
for $M_H=130\,\gev$. The dashed line shows the total background
contribution whereas the dotted line corresponds to the real
$\gamma\gamma$ background. The solid line displays the expected
contribution of signal plus background.

In Table~\ref{tab:results} results are given in terms of the total
number of expected signal events, $S$, and background events, $B$,
for 30 fb$^{-1}$ of accumulated luminosity. The signal
significance is given with the Gaussian approximation,
$S/\sqrt{B}$, and the single bin Poisson calculation.

%As pointed out in Section~\ref{sec:doublecounting}, the estimation
%of the $\gamma\gamma jj$ background made here is rather
%conservative

The QCD $\gamma\gamma jj$ has
been estimated with QCD $\gamma\gamma jj$ ME based MC alone (see Section~\ref{sec:doublecounting}). The rate of additional (non tagging) jets has been
estimated with the help of the parton shower. This approach yields
a central jet veto survival probability significantly smaller than
that calculated in~\cite{Rainwaterthesis}. It should not be
forgotten that this feature is present in all the analyses of the
VBF modes made public so far by the ATLAS collaboration. Both
effects go in the direction of the overestimation of the
$\gamma\gamma jj$ background. Similar discussion applies to the estimation of the fake photon background performed here.

\begin{figure}[ht]
{\centerline{\epsfig{figure=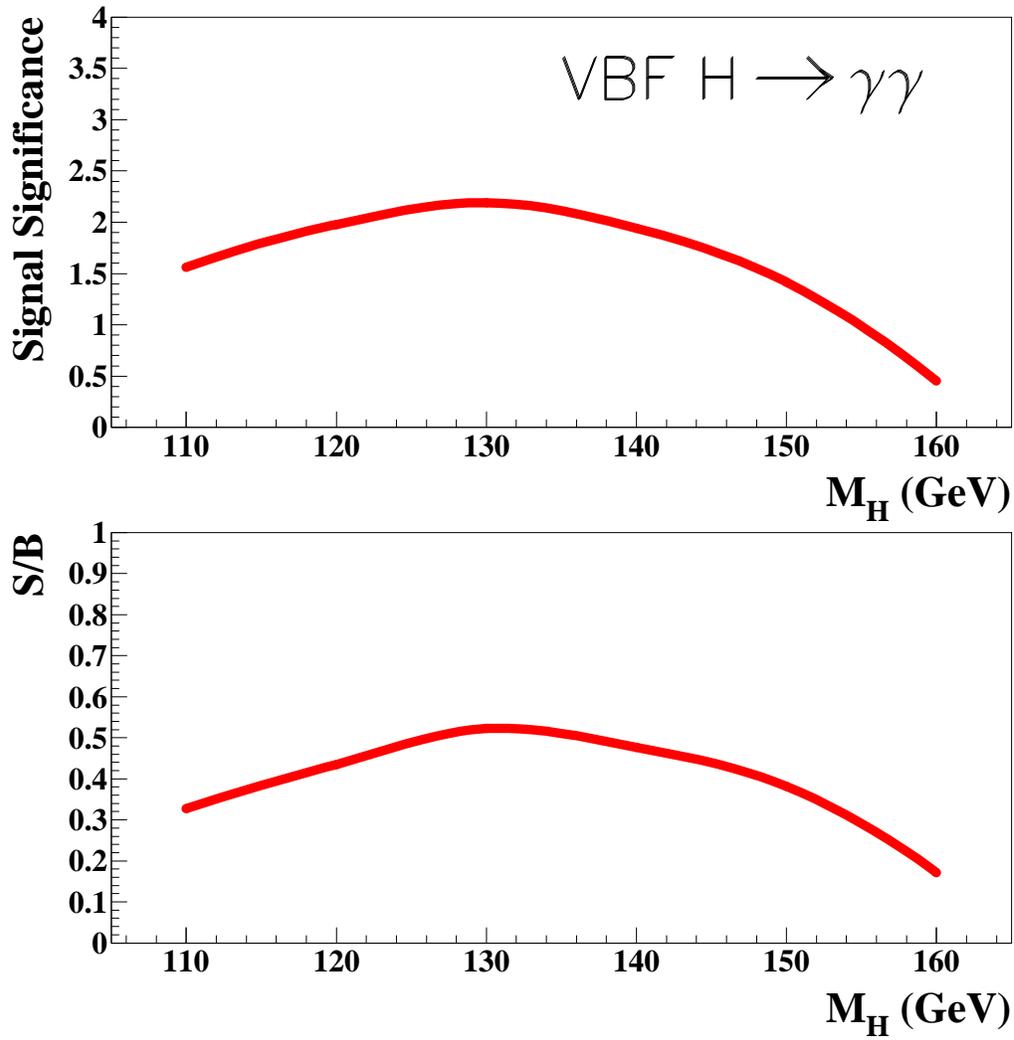,width=14cm}}}
\caption[]{The upper plot displays the expected signal significance as a function of the Higgs mass for 30~fb$^{-1}$ of accumulated luminosity. The lower plot shows the expected ratio of signal to background as a function of the Higgs mass.} \label{fig:results}
\end{figure}

In conclusion, the signal significance expected with this VBF mode
alone reaches up to 2.2$\,\sigma$ for 30 fb$^{-1}$ of accumulated
luminosity. These results are summarized in
Figure~\ref{fig:results}. The upper and lower plots in
Figure~\ref{fig:results} display the signal significance (for 30
fb$^{-1}$ of accumulated luminosity) and signal to background
ratio dependencies on the Higgs mass.

This estimation may be improved with the implementation of a more
realistic MC for the simulation of the real photon background. A
better understanding of fake photon rejection would significantly
help this analysis, as well.

 \begin{table}[ht]
 \begin{center}
 \begin{tabular}{||c||c|c||c|c||c|c||}
 \hline
 Cut & VBF H & g-g Fusion H & QCD $\gamma\gamma jj$ &      EW $\gamma\gamma jj$ & $\gamma jjj$ & $jjjj$ \\
\hline\hline
 {\bf a} &       2.25 &       5.45 &     246.90 &       7.97 &     172.60 &     691.06  \\
 \hline
 {\bf b} &       0.73 &       0.08 &      31.83 &       4.39 &      28.30 &      35.22  \\
 \hline
 {\bf c} &       0.70 &       0.07 &      16.81 &       4.20 &      21.76 &      30.06  \\
 \hline
 {\bf d} &       0.57 &       0.04 &       7.43 &       3.69 &      12.77 &      16.99  \\
 \hline
 {\bf e} &       0.42 &       0.02 &       5.41 &       2.50 &       8.52 &       8.49  \\
 \hline
 {\bf f} &       0.38 &       0.02 &       0.28 &       0.14 &       0.22 &       0.25  \\
 \hline
 \end{tabular}
 \caption{Expected signal and background                     cross-sections (in fb) corrected for acceptance                   and efficiency corrections after the application of               successive cuts. Here $M_H=120\,\gev$.}
 \label{tab:cross120}
 \end{center}
 \end{table}

 \begin{table}[ht]
 \begin{center}
 \begin{tabular}{||c||c|c||c|c||c|c||}
 \hline
 $M_H$ & VBF H & g-g Fusion H & QCD $\gamma\gamma jj$ &    EW $\gamma\gamma jj$ & $\gamma jjj$ & $jjjj$ \\
\hline\hline
  110 &       0.32 &       0.02 &       0.29 &       0.14 &       0.25 &       0.35  \\
 \hline
  120 &       0.38 &       0.02 &       0.28 &       0.14 &       0.22 &       0.25  \\
 \hline
  130 &       0.39 &       0.02 &       0.26 &       0.13 &       0.21 &       0.20  \\
 \hline
  140 &       0.34 &       0.02 &       0.25 &       0.11 &       0.21 &       0.20  \\
 \hline
  150 &       0.24 &       0.01 &       0.22 &       0.10 &       0.17 &       0.18  \\
 \hline
  160 &       0.09 &       0.01 &       0.18 &       0.08 &       0.13 &       0.18  \\
 \hline
 \end{tabular}
 \caption{Expected signal and background                    cross-sections, in fb, corrected for acceptance                   and efficiency corrections after the application of all cuts. Cross-sections are given as a function of $M_H$.}
 \label{tab:cross}
 \end{center}
 \end{table}

 \begin{table}[ht]
 \begin{center}
 \begin{tabular}{||c|c|c|c|c|c||}
 \hline
 $M_H$ & $S$ & $B$ & $S/B$ & $S/\sqrt{B}$ & $\sigma_P$     \\
\hline\hline
  110 &      10.05 &      30.69 &       0.33 &       1.82 &       1.56  \\
 \hline
  120 &      12.06 &      26.54 &       0.45 &       2.34 &       2.02  \\
 \hline
  130 &      12.52 &      23.97 &       0.52 &       2.56 &       2.19  \\
 \hline
  140 &      10.91 &      22.90 &       0.48 &       2.28 &       1.94  \\
 \hline
  150 &       7.69 &      20.15 &       0.38 &       1.71 &       1.42  \\
 \hline
  160 &       2.89 &      17.21 &       0.17 &       0.70 &       0.44  \\
 \hline
 \end{tabular}
 \caption{Expected number of signal and background          events and the corresponding signal significance                  for 30~fb$^{-1}$ of accumulated luminosity. }
 \label{tab:results}
 \end{center}
 \end{table}

\bibliographystyle{zeusstylem}
\bibliography{vbf,mycites}

\end{document}